\documentclass[mypaper,7pt,twoside]{CoAst}
\usepackage{epsf,graphicx,fancyhdr}
\renewcommand{\chaptermark}[1]{\markboth{#1}{}}

\def\kms {km~s$^{-1}$}

\newcommand{\kopf}{\small\itshape Comm. in Asteroseismology \\ Contribution to the Proceedings of the Wroclaw HELAS Workshop, 2008}

\newcommand{\Authors}[1]{\begin{center}\normalsize\bf\sf #1 \end{center}}

\renewcommand{\author}[1]{\begin{center}\normalsize\bf\sf #1 \end{center}}
\newcommand{\Address}[1]{\begin{center}\small\sf #1 \end{center}}

\newcommand{\Session}[1]{{\vspace{3mm}\small \noindent  \hspace*{3mm} Session: } #1 \normalsize}

	\newcommand{\threeD}{\small STARS - effects of magnetic field on stellar pulsation \newline}

\renewenvironment{abstract}{\section*{Abstract}\normalsize\sf}{}
\newcommand{\References}[1]{\begin{flushleft}{\large References\\}\vspace*{2mm}\small #1 \end{flushleft}}

\newcommand{\chapterCoAst}[2]{\chapter[\sf\normalsize #1\\ \footnotesize \hspace*{5mm}by #2 \sf\normalsize][]{#1\\}\rhead[\fancyplain{}{\sf\footnotesize \center{#1}}]{\fancyplain{}{\sffamily\thepage}}\lhead[\fancyplain{\kopf}{\sffamily\thepage}]{\fancyplain{\kopf}{\sf\footnotesize \center{#2}}}}

\renewcommand{\chaptername}{}

\def\rfr{\smallskip\par\noindent
        \hangindent=7truemm
        \hangafter=1}

\def\ms{m\,s$^{-1}$}

\begin{document}
\sf

\chapterCoAst{Pulsation in the atmosphere of roAp stars}
{O.\,Kochukhov} 
\Authors{O.\,Kochukhov} 
\Address{Department of Physics and Astronomy, Uppsala University, \\Box 515,SE-751 20 Uppsala, Sweden}

\noindent
\begin{abstract}
High time resolution spectroscopy of roAp stars at
large telescopes has led to a major breakthrough in our understanding of
magnetoacoustic pulsations in these interesting objects. New observations have
allowed to uncover a number of intricate relations between stellar
oscillations, magnetic field, and chemical inhomogeneities. 
It is now understood that unusual pulsational
characteristics of roAp stars arise from an interplay between short vertical
length of pulsation waves and extreme chemical stratification. Here I review
results of recent studies which utilize these unique properties 
to map 3D pulsation geometry using a combination of Doppler imaging,
vertical pulsation tomography, interpretation of line profile variation, and
ultraprecise space photometry. I also describe recent attempts to interpret
theoretically the complex observational picture of roAp pulsations.
\end{abstract}

\Session{\threeD}

\section*{Introduction}

Rapidly oscillating Ap (roAp) stars represent an interesting subgroup of chemically peculiar
(SrCrEu type)
magnetic A stars pulsating in high-overtone, low degree {\it p-}modes. These stars are found
at or near the main sequence, close to the cool border of the region occupied by the magnetic Ap/Bp
stars (Kochukhov \& Bagnulo 2006). According to the series of recent spectroscopic  studies
(e.g., Kochukhov et al. 2002; Ryabchikova et al. 2004), effective
temperatures of roAp stars range from about 8100 down to 6400~K. Their atmospheres are
characterized by diverse chemical abundance patterns, but typically have normal or
below solar concentration of light and iron-peak elements and a very large overabundance of rare-earth
elements (REEs). Similar to other cool magnetic A  stars, roAp stars possess global fields
with a typical strength of a few kG (Mathys et al. 1997), although in some stars the field
intensity can exceed 20~kG (Kurtz et al. 2006b). These global magnetic topologies are most
likely the remnants of the fields which were swept at the star-formation phase or generated
by dynamo in convective pre-main sequence stars, decayed to a stable
configuration (Braithwaite \& Nordlund 2006) and now remain nearly constant on long
timescales. The slow rotation and stabilizing effect of the strong magnetic field
facilitates operation of atomic diffusion (Michaud et al. 1981; LeBlanc \& Monin 2004),
which is responsible for the  grossly non-solar surface chemistry and large element concentration
gradients in Ap-star atmospheres (Ryabchikova et al. 2002, 2008; Kochukhov et al. 2006).
Variation of the field strength and inclination across the stellar surface alters the local
diffusion velocities (Alecian \& Stift 2006), leading to the formation of spotted chemical
distributions and consequential synchronous rotational modulation of the broad-band
photometric indices, spectral line profiles, longitudinal magnetic field and mean field
modulus (e.g., Ryabchikova et al. 1997).

Pulsations in cool Ap stars were discovered 30 years ago (Kurtz 1978) and were
immediately recognized to be another manifestation of the prominent influence of  unusually
strong magnetic fields on the stellar interiors and atmospheres.  Currently (mid 2008), 40
cool Ap stars are known to pulsate, with several new roAp stars discovered by high-resolution
spectroscopic observations (Hatzes \& Mkrtichian 2004; Elkin et al. 2005; Kurtz et al. 2006b;
Kochukhov et al. 2008a, 2008b; Gonz\'ales et al. 2008). Oscillations have amplitudes below 10
mmag in the Johnson's B filter and 0.05--5~\kms\ in spectroscopy, while the periods lie in
the range from 4 to 22\footnote{The longest roAp pulsation period corresponds to the second
mode recently detected by high-precision HARPS observations  of the evolved Ap star
HD\,116114 (Kochukhov, Bagnulo \& Lo Curto, in preparation).} min. The amplitude and phase of
pulsational variability are modulated with the stellar rotation. A simple geometrical
interpretation of this phenomenon was suggested by the oblique pulsator model of Kurtz (1982),
which supposes an alignment of the low angular degree modes with the quasi-dipolar magnetic
field of the star and resulting variation of the aspect at which pulsations are seen by the
distant observer. Recent theoretical studies (Bigot \& Dziembowski 2002; Saio 2005) indicated
that the horizontal pulsation picture of {\it p-}mode pulsations in magnetic stars is far
more complicated: individual modes are distorted by the magnetic field and rotation in such a
way  that pulsational perturbation cannot be approximated by a single spherical harmonic
function.

The question of the roAp excitation mechanism has been debated for many years but now is
narrowed down to the $\kappa$ mechanism acting in the H~{\sc i} ionization zone with the
additional influence from the magnetic quenching of convection and composition gradients
built up by the atomic diffusion (Balmforth et al. 2001; Cunha 2002; Vauclair \& Th\'eado
et al. 2004). However, theories cannot reproduce the observed temperature and
luminosity distribution of roAp stars and have not been able to identify parameters
distinguishing pulsating Ap stars from their apparently constant, but otherwise very similar,
counterparts. On the other hand, impressive success has been achieved in calculating magnetic
perturbation of oscillation frequencies (Cunha \& Gough 2000; Saio \& Gautschy 2004) and inferring fundamental
parameters and  interior properties for multiperiodic roAp stars (Cunha et al. 2003;
Gruberbauer et al. 2008; Huber et al. 2008).

\section*{Photometric studies of roAp pulsations}

Majority of roAp stars were discovered by D. Kurtz and collaborators using photometric
observations at SAAO (see review by Kurtz \& Martinez 2000). Few roAp stars were also
observed in coordinated multi-site  photometric campaigns (Kurtz et al. 2005a), which allowed
to deduce frequencies with the precision sufficient for asteroseismic analysis. However, low
amplitudes of broad-band photometric variation of roAp stars, low duty cycle and aliasing
problems inevitably limit precision of the ground-based photometry. Instead of pursuing
observations from the ground, recent significant
progress has been achieved by uninterrupted, ultra-high precision observations of known roAp
stars using small photometric telescopes in space. Here the Canadian MOST space telescope
is undisputed leader. The MOST team has completed 3--4 week runs on HR~24712, $\gamma$~Equ,
10~Aql,  HD~134214, and HD~99563. 

Asteroseismic interpretation of the frequencies deduced
from the  MOST data for $\gamma$~Equ (Gruberbauer et al. 2008) and 10~Aql (Huber et al. 2008)
yields stellar parameters in good agreement with those determined in detailed model
atmosphere studies. At the same time, magnetic field required by the seismic models to
fit observed frequencies is 2--3 times stronger than the field modulus inferred from the Zeeman
split spectral lines. This discrepancy could be an indication that magnetic field in the
{\it p-}mode driving zone is significantly stronger than the surface field or it may 
reflect limitations of theoretical models.

MOST photometry of $\gamma$~Equ has also revealed the presence of a very close frequency pair
giving modulation of pulsation amplitude with $\approx$18~d period (Huber et al. 2008). It is
possible that this frequency beating is responsible for significant discrepancy of radial
velocity amplitudes found for $\gamma$~Equ in different spectroscopic observing runs (Sachkov
et al., this meeting). This amplitude variation could not be ascribed to the rotational
modulation because rotation period of this star exceeds 70 years (Bychkov et al. 2006).

\section*{Spectroscopy of roAp pulsations}

High-quality time-resolved spectra of roAp stars have proven to be the
source of new, incredibly rich information, which not only opened new possibilities for the
research on magnetoacoustic pulsations but yielded results of wide astrophysical
significance. Numerous spectroscopic studies of individual roAp stars (e.g., Kochukhov \&
Ryabchikova 2001a; Mkrtichian et al. 2003; Ryabchikova et al. 2007a), as well as comprehensive
analysis of pulsational variability in 10 roAp stars published by Ryabchikova et al. (2007b),
demonstrated pulsations in spectral lines very different from those observed in any other type  of
non-radially pulsating stars. The most prominent characteristic of the RV oscillation in roAp
stars is the extreme diversity of pulsation signatures seen in the lines of different
elements. Only a few stars show evidence of $<$50~\ms\ variation in the lines of iron-peak
elements, whereas REE lines, especially those of Nd~{\sc ii}, Nd~{\sc iii}, Pr~{\sc iii} and
Dy~{\sc iii}, exhibit amplitudes from a few hundred \ms\ to several \kms. The narrow core of
H$\alpha$ behaves similarly to REE lines (Kochukhov 2003; Ryabchikova et
al. 2007b), suggesting line formation at comparable atmospheric heights.

Pulsation phase also changes significantly from one line to another (Kochukhov \& Ryabchikova
2001a; Mkrtichian et al. 2003), with the most notorious example of 33~Lib where different
lines of {\it the same ion} pulsate with a 180$^{\rm o}$ shift in phase, revealing a radial
node, and show very different ratios of the amplitude at the main frequency and its
first harmonic (Ryabchikova et al. 2007b). Several studies concluded that, in general, roAp
stars show a combination of running (changing phase) and
standing (constant phase) pulsation wave behaviour at different atmospheric heights.

Another unusual aspect of the spectroscopic pulsations in roAp stars is a large change of
oscillation amplitude and phase from the line core to the wings. Bisector variation expected
for the regular spherical harmonic oscillation is unremarkable and should exhibit neither
changing phase nor significantly varying amplitude. Contrary to this expectation of the
common single-layer pulsation model, roAp bisector 
amplitude often shows an increase from 200--400~\ms\ in the cores of strong REE lines to
2--3~\kms\ in the line wings, accompanied by significant changes of bisector phase (Sachkov
et al. 2004; Kurtz et al. 2005b; Ryabchikova et al. 2007b).

The ability to resolve and measure with high precision pulsational variation in individual lines
allows to focus analysis on the spectral features most sensitive to pulsations. By co-adding radial
velocity curves of many REE lines one is able to reach the RV accuracy of $\sim1$~\ms. This
made possible discovery of the low-amplitude oscillations in HD\,75445 (Kochukhov et al.
2008b) and HD\,137909 (Hatzes \& Mkrtichian 2004). The second object, well-known cool Ap star
$\beta$~CrB, was previously considered to be a typical non-pulsating Ap (noAp) star due to 
null results of numerous photometric searches of pulsations (Martinez \& Kurtz 1994) and the
absence of prominent REE ionization anomaly found for nearly all other roAp stars (Ryabchikova et
al. 2001, 2004). The fact that $\beta$~CrB is now revealed as the second brightest roAp star
corroborates the idea that {\it p-}mode oscillations could be present in all cool Ap stars
but low pulsation amplitudes prevented detection of pulsations in the so-called noAp stars
(Ryabchikova et al. 2004).

Despite improved sensitivity in searches of the low-amplitude oscillations in roAp candidates
and numerous outstanding discoveries for known roAp stars, the major drawback of the
high-resolution spectroscopic monitoring is still a relatively small amount of observing time
available at large telescopes for these projects. As a result, only short time-series
spanning 2--4 hours were recorded for most roAp stars, thus providing an incomplete picture
for multiperiodic pulsators where different frequencies cannot be resolved in such short
runs. Observations on different nights required to infer detailed RV frequency spectrum were
secured only for a few roAp stars (Mkrtichian \& Hatzes 2005, Kochukhov 2006). In recent
multi-site spectroscopic campaign carried out for 10~Aql using two telescopes on 7 different
observing nights (Sachkov et al. 2008), we found that beating of the three dominant
frequencies leads to strong changes of the apparent RV amplitude during several hours. This
phenomenon could explain puzzling modulation of RV pulsations on the timescale of 1--2 hours
detected in  some roAp stars (Kochukhov \& Ryabchikova 2001b; Kurtz et al. 2006a).

\section*{Interpretation of roAp oscillations}

The key observational signature of roAp pulsations in spectroscopy -- large line-to-line
variation of pulsation amplitude and phase -- is understood in terms of an interplay between
pulsations and chemical stratification. The studies by Ryabchikova et al. (2002, 2008) and
Kochukhov et al. (2006) demonstrated that light and iron-peak elements tend to be
overabundant in deep atmospheric layers (typically $\log\tau_{5000}\ge -0.5$) of cool Ap
stars, which agrees with the predictions of self-consistent diffusion models (LeBlanc \&
Monin 2004). On the other hand, REEs accumulate in a cloud located above
$\log\tau_{5000}\approx -3$ (Mashonkina et al. 2005). Then, the rise of pulsation amplitude
towards the upper atmospheric layers due to exponential density decrease does not affect Ca,
Fe, and Cr lines but shows up prominently in the core of H$\alpha$ and in REE lines. This
picture of the pulsation waves propagating outwards through the stellar atmosphere with
highly inhomogeneous chemistry has gained general support from observations and theoretical
studies alike. Hence the properties of roAp atmospheres allow an entirely new type of
asteroseismic analysis -- vertical resolution of {\it p-}mode cross-sections simultaneously
with the constraints on distribution of chemical abundances. 

The two complimentary approaches to the pulsation tomography problem have been discussed by
Ryabchikova et al. (2007a, 2007b). On the one hand, tedious and detailed line formation
calculations, including stratification analysis, NLTE line formation, sophisticated model
atmospheres and polarized radiative transfer, can supply mean formation heights for
individual pulsating lines. Then, the pulsation mode structure can be mapped directly by
plotting pulsation amplitude and phase of selected lines against optical or geometrical depth.
On the other hand, the phase-amplitude diagram method proposed by Ryabchikova et al. (2007b)
is suitable for a coarse analysis of the vertical pulsation structure without invoking model
atmosphere calculations but assuming the presence of the outwardly propagating wave
characterized by continuous change of amplitude and phase. In this case, a scatter plot of the
RV measurements in the phase-amplitude plane can be interpreted in terms of the standing and
running waves, propagating in different parts of the atmosphere.

To learn about the physics of roAp atmospheric oscillations one should compare
empirical pulsation maps with theoretical models of the {\it p-}mode
propagation in magnetically-dominant ($\beta<<1$) part of the stellar envelope. Sousa \& Cunha (2008)
considered an analytical model of the radial modes in an isothermal atmosphere with
exponential density decrease. They argue that waves are decoupled into the standing magnetic
and running acoustic components, oriented perpendicular and along magnetic field lines,
respectively. The total projected pulsation velocity, produced by a superposition of these
two components, can have widely different vertical profile depending on the magnetic field
strength, inclination and the aspect angle. For certain magnetic field parameters and viewing
geometries the two  components cancel out, creating a node-like structure. This model
can possibly account for observations of radial nodes in 33~Lib (Mkrtichian et al.
2003)  and 10~Aql (Sachkov et al. 2008). 

The question of interpreting the line profile variation (LPV) of roAp stars has recieved
great attention after it was demonstrated that REE lines in $\gamma$~Equ exhibit
unusual blue-to-red asymmetric variation (Kochukhov \& Ryabchikova 2001a), which is entirely 
unexpected for a slowly
rotating non-radial pulsator. Kochukhov et al. (2007) showed the presence of similar LPV in
the REE lines of many other roAp stars and presented examples of the transformation from the
usual symmetric blue-red-blue LPV in Nd~{\sc ii} lines to the asymmetric blue-to-red waves in
the Pr~{\sc iii} and Dy~{\sc iii} lines formed higher in the atmosphere. These lines often
show anomalously broad profiles (e.g., Ryabchikova et al. 2007b), suggesting existence of
an isotropic velocity field of the order of 10~\kms\ in the uppermost atmospheric layers. Kochukhov
et al. (2007) proposed a model of the interaction between this turbulent layer and
pulsations that has successfully reproduced asymmetric LPV of doubly ionized REE lines. An
alternative model by Shibahashi et al. (2008) obtains similar LPV by postulating formation
of REE lines at extremely low optical depths, in disagreement with the detailed NLTE
calculations by Mashonkina et al. (2005), and requires the presence of shock waves in stellar
atmospheres, which is impossible to reconcile with the fact that observed RV amplitudes are
well below the sound speed.

Oblique pulsations and distortion of modes by rotation and magnetic field precludes
application of the standard mode identification techniques to roAp stars. A meaningful study
of their horizontal pulsation geometry became possible by using the method of pulsation
Doppler imaging (Kochukhov 2004a). This technique derives maps of pulsational fluctuations
without making {\it a priori} assumption of the spherical harmonic pulsation geometry.
Application of this method to HR\,3831 (Kochukhov 2004b) provided the first independent proof
of the oblique pulsator model by showing alignment of the axisymmetric pulsations with
magnetic field. At the same time, Saio (2005) showed that the observed deviation of the
oscillation geometry of HR\,3831 from a oblique dipole mode agrees well with his model
of magnetically distorted pulsation.

\References{
\rfr Alecian G., \& Stift M.J. 2006, A\&A, 454, 571
\rfr Balmforth N.J., Cunha M.S, Dolez N., et al. 2001, MNRAS, 323, 362
\rfr Bigot L., \& Dziembowski W.A. 2002, A\&A, 391, 235
\rfr Braithwaite J., \& Nordlund \AA. 2006, A\&A, 450, 1077
\rfr Bychkov V.D., Bychkova L.V., \& Madej J. 2006, MNRAS, 365, 585
\rfr Cunha M.S., \& Gough D. 2000, MNRAS, 319, 1020
\rfr Cunha M.S. 2002, MNRAS, 333, 47
\rfr Cunha M.S., Fernandes J.M.M.B., \& Monteiro, M.J.P.F.G. 2003, MNRAS, 343, 831  
\rfr Elkin V.G., Rilej J., Cunha M., et al. 2005, MNRAS, 358, 665
\rfr Gonz\'ales J.F., Hubrig S., Kurtz D.W., et al. 2008, MNRAS, 384, 1140
\rfr Gruberbauer M., Saio H., Huber D., et al. 2008, A\&A, 480, 223
\rfr Hatzes A.P., \& Mkrtichian D.E. 2004, MNRAS, 351, 663
\rfr Huber D., Saio H., Gruberbauer M., et al. 2008, A\&A, 483, 239
\rfr Kochukhov O., \& Ryabchikova T. 2001a, A\&A, 374, 615
\rfr Kochukhov O., \& Ryabchikova T. 2001b, A\&A, 377, L22
\rfr Kochukhov O. 2003, in {\it Magnetic Fields in O, B and A stars}, eds.
Balona L.A., Henrichs H.F., \& Medupe R., ASP Conf. Ser., 305, 104
\rfr Kochukhov O. 2004a, A\&A, 423, 613
\rfr Kochukhov O. 2004b, ApJ, 615, L149
\rfr Kochukhov O. 2006, A\&A, 446, 1051
\rfr Kochukhov O., \& Bagnulo S. 2006, A\&A, 450, 763
\rfr Kochukhov O., Bagnulo S., \& Barklem P.S. 2002, ApJ, 578, L75
\rfr Kochukhov O., Tsymbal V., Ryabchikova T., et al. 2006, A\&A, 460, 831
\rfr Kochukhov O., Ryabchikova T., Weiss W.W., et al. 2007, MNRAS, 376, 651
\rfr Kochukhov O., Ryabchikova T., Bagnulo S., \& Lo Curto G. 2008a, A\&A, 479, L29
\rfr Kochukhov O., Ryabchikova T., Bagnulo S., \& Lo Curto G. 2008b, CoSka, 38, 423
\rfr Kurtz D.W. 1978, IBVS, 1436
\rfr Kurtz D.W. 1982, MNRAS, 200, 807
\rfr Kurtz D.W., \& Martinez, P. 2000, Baltic Astronomy, 9, 253
\rfr Kurtz D.W., Elkin V.G., \& Mathys G. 2005a, MNRAS, 358, L10
\rfr Kurtz D.W., Cameron C., Cunha M.S., et al. 2005b, MNRAS, 358, 651
\rfr Kurtz D.W., Elkin V.G., \& Mathys G. 2006a, MNRAS, 370, 1274
\rfr Kurtz D.W., Elkin V.G., Cunha M.S., et al. 2006b, MNRAS, 372, 286
\rfr LeBlanc F., \& Monin D. 2004, in {\it IAU Symposium 224},
eds. {Zverko} J., {Ziznovsky} J., {Adelman} S.~J., \& {Weiss} W.~W., 193
\rfr Mashonkina L., Ryabchikova T., \& Ryabtsev V. 2005, A\&A, 441, 309
\rfr Mathys G., Hubrig S., Landstreet J.D. et al. 1997, A\&AS, 123, 353
\rfr Martinez P., \& Kurtz D.W. 1994, MNRAS, 271, 129
\rfr Michaud G., Charland Y., \& Megessier C. 1981, A\&A, 103, 244
\rfr Mkrtichian D.E., Hatzes A.P., \& Kanaan A. 2003, MNRAS, 345, 781
\rfr Mkrtichian D.E., \& Hatzes A.P. 2005, JApA, 26, 185
\rfr Ryabchikova T.A., Landstreet J.D., Gelbmann M.J., et al. 1997, A\&A, 327, 1137
\rfr Ryabchikova T.A., Savanov I.S., Malanushenko V.P., Kudryavtsev D.O. 2001, Astron.
Reports, 45, 382
\rfr Ryabchikova T., Piskunov N., Kochukhov O., et al. 2002, A\&A, 384, 545
\rfr Ryabchikova T., Nesvacil N., Weiss W.W., et al. 2004, A\&A, 423, 705
\rfr Ryabchikova T., Sachkov M., Weiss W.W., et al. 2007a, A\&A, 462, 1103
\rfr Ryabchikova T., Sachkov M., Kochukhov O., \& Lyashko D. 2007b, A\&A, 473, 907
\rfr Ryabchikova T., Kochukhov O., \& Bagnulo S. 2008, A\&A, 480, 811
\rfr Sachkov M., Ryabchikova T., Kochukhov O., et al. 2004, in {\it IAU Colloquium 193},
eds. Kurtz D.W., \& Pollard K.R., ASP Conf. Ser., 310, 208
\rfr Sachkov M., Kochukhov O., Ryabchikova T., et al. 2008, MNRAS, 389, 903
\rfr Saio H., \& Gautschy A. 2004, MNRAS, 350, 485
\rfr Saio H. 2005, MNRAS, 360, 1022
\rfr Shibahashi H., Gough D., Kurtz D.W., \& Kambe E. 2008, PASJ, 60, 63
\rfr Sousa J., \& Cunha, M.S. 2008, CoSka, 38, 453
\rfr Vauclair S., \& Th\'eado S. 2004, A\&A, 425, 179
}

\end{document}